\documentclass[prb,a4paper,twocolumn,showpacs]{revtex4}

\usepackage{graphicx}
\usepackage{bm}

\def\w{\omega}

\begin{document}

\title{On  X-ray-singularities in the  f-electron spectral function of the 
Falicov-Kimball model}

\smallskip

\author{Frithjof B.~Anders}
\affiliation{Department of Physics, Universit\"at Bremen,
                  P.O. Box 330 440, D-28334 Bremen, Germany}
\author{G.~Czycholl}
\affiliation{Department of Physics, Universit\"at Bremen,
                  P.O. Box 330 440, D-28334 Bremen, Germany}

\date{November 22, 2004}

\begin{abstract}
The f-electron spectral function of the Falicov-Kimball model is calculated
within the dynamical mean-field theory using the numerical renormalization
group method as the impurity solver. Both the Bethe lattice and the
hypercubic lattice are considered at half filling. For small U 
we obtain a single-peaked f-electron spectral function, which --for zero
temperature-- exhibits  an algebraic
(X-ray) singularity ($|\omega|^{-\alpha}$) for   $\omega \rightarrow 0$. The
characteristic exponent $\alpha$ depends on the Coulomb (Hubbard)
correlation U. This X-ray singularity
cannot be observed when using alternative (Keldysh-based) many-body 
approaches. With increasing U, $\alpha$ decreases and vanishes for
sufficiently large U when the f-electron spectral function develops a gap
and a two-peak structure (metal-insulator transition).
 
\end{abstract}

\maketitle

\def\e{\varepsilon}
\def\as{\alpha\sigma}
\def\ket#1{|#1\rangle}
\def\bra#1{\langle #1|}
\def\expect#1{\langle #1\rangle}
\def\matrixElement#1#2#3{\langle #1|#2| #3\rangle}
\def\green#1#2{\ll \!\! #1| #2 \!\! \gg}
\def\sus{susceptibility}
\newcommand{\punkt}{\;\; .}
\newcommand{\komma}{\;\; ,}

\section{Introduction:}

One of the simplest lattice models for strongly correlated electron
systems, the  Falicov-Kimball model (FKM) \cite{FalicovKimball69}, consists
of two types of spinless electrons, delocalized band 
($c$-) electrons, localized $f$-electrons and a local Coulomb (Hubbard)
interaction between $c$- and $f$-electron at the same site. 
The FKM Hamiltonian reads
\begin{eqnarray}
  \label{eqn:fkm-h}
  H &=& -\sum_{ij} t_{ij} c^\dagger_i c_j + \e_c\sum_i c^\dagger_i c_i
+
\e_f 
\sum_i f^\dagger_i  f_i 
\nonumber
\\
&&
+U
\sum_i  c^\dagger_i c_i f^\dagger_i  f_i 
\punkt
\end{eqnarray}
where $\e_f$ denotes the one-particle energy of the $f$-electrons, $\e_c$ the
band center of the $c$-electrons, $t_{i j}$ the c-electron hopping between
site $i$ and $j$ and U the Coulomb (Hubbard) correlation.
This model can  be interpreted as a simplified
version of the one-band 
Hubbard model \cite{Hubbard64} where $c$ and $f$ translates to the
label of the different spin orientations, and the hopping term for the
$f$-spin type is set to zero. The FKM was originally introduced as a model
for metal-insulator and valence transitions \cite{FalicovKimball69}. It can
also be interpreted as a model for crystallization (identifying the
''heavy'' $f$-particles with the  nuclei, the $c$-particles with the
electrons and for $U<0$) \cite{KennedyLieb86}. Furthermore, the FKM
is of interest for 
academic reasons, because it is the simplest, non-trivial lattice model for
correlated electron systems, for which certain exact results are available;
for a recent review see Ref. \onlinecite{FreericksZlatic03}. Therefore, the
FKM may serve as a test case and benchmark for any many-body method and
approximation, because such a method can also be applied to the FKM and
comparisons of the approximate and the available exact results may help to
judge the value and possible limitations of such  many-body techniques.


Brandt and Mielsch \cite{BrandtMielsch89} solved the FKM exactly in
the limit of infinite spatial dimensions, 
$d \rightarrow \infty$ \cite{MetznerVollhardt89},
 by  mapping the problem onto an effective
single site subject to a complex bath. Thereby, the complex bath field
mimics the  dynamics of the surrounding electrons on the lattice. Since
the local $f$-electron number is conserved, $[H,f^\dagger_i f_i]=0$,
the $c$-electron problem is solved independently from the dynamics of the
$f$-electrons which are static. However, the presence or absence of an
$f$-electron on the effective site induces a significant change for the
conduction electrons: the presence of an $f$-electron switches on an
additional scattering potential $U$ for the conduction
electrons. Brandt and Mielsch\cite{BrandtMielsch89}
 were able to derive an exact, explicit
 expression for the self-energy of the $c$-electrons
$\Sigma_c(z)$, which is a functional of the local $c$-electron Green function
and the (fixed) local $f$-electron occupation numbers. This functional is
essentially equivalent to that obtained in the Hubbard-III-approximation of
the Hubbard model\cite{Hubbard64}, i.e. this approximation becomes exact for
the $c$-electron self-energy of the FKM in the limit $d \rightarrow \infty$.

But despite the conservation of the $f$-electron occupation, the
$f$-electrons have a highly non-trivial dynamics. For $U=0$ the f-electron
spectral function is simply a delta-function, but for finite $U$ the
spectral function broadens due to the interaction with the fluctuating
$c$-electrons. The 
$f$-electron Green function $G_f(z)$ of the FKM is much more involved and less
trivial than the $c$-electron Green function $G_c$, in particular
 a Hubbard-III-type
self-energy functional does not become exact for the $f$-electrons. 
Even for large dimensions, where the mapping on an effective
impurity problem holds, it 
is not possible to derive an explicit analytical expression for
the $f$-electron self-energy functional $\Sigma_f$. On the contrary, the
evaluation of $\Sigma_f$ is nearly as complicated as  the
self-energy of the full Hubbard model. For $d \rightarrow \infty$, the first
explicit numerical evaluation of $G_f$ (or the $f$-electron spectral function
$\rho_f$, respectively) was performed by Brandt and
Urbanek\cite{BrandtUrbanek92}. They started from a nonequilibrium, 
Keldysh based many-body formalism using a discretization along the
Kadanoff-Baym contour and a subtle analytical continuation to obtain the
dependence on real frequency numerically. Only relatively few explicit
results were presented in Ref. \onlinecite{BrandtUrbanek92}. Therefore,
this Keldysh based  analysis was repeated very recently by 
Freericks et al.\cite{FreericksZlatic2004} presenting much more explicit
results and investigating also the accuracy
of the results by checking sum rules and 
the dependence on the numerical discretization
and the cutoff in the time integrals.  

On the other hand, it is clear that the problem to calculate the 
$f$-electron Green function
$G_{f}(\tau) = - \expect{T(f_i(\tau) f^\dagger_i)}$  must be related to
the X-ray threshold problem, because the creation of a core hole, with which
the conduction electrons then interact, is related to the creation of an
f-electron at a certain time as pointed out  by Si {\em
  et~al.~}\cite{SiKotliarGeorges1992} and M\"oller {\em
  et~al.~}\cite{MoellerRuckensteinSchmittRink1992} As in the X-ray problem
 the non-interacting Fermi sea  of $c$-electrons
has to relax into an orthogonal ground state through an infinite
cascade of particle-hole excitations after an $f$-electron has been added
to the system by $ f^\dagger_i$. However, though the relation to the X-ray
problem is known, no observations of the characteristic X-ray singularities
were made in previous studies of $G_f$ for the FKM.
\cite{BrandtUrbanek92,FreericksZlatic2004}  

In this paper, we apply a different method to the FKM, namely the dynamical
mean-field theory (DMFT), i.e. the (exact) mapping of the 
$d \rightarrow \infty$ problem on an effective impurity 
problem\cite{BrandtMielsch89,Georges96}, and Wilson's numerical
renormalization group approach (NRG)\cite{Wilson75} as the impurity solver.
Within this DMFT/NRG  the typical X-ray singularities, in fact,
occur for $G_f$ of the FKM. The spectral function of
$G_f$ exhibits a power law $|\w|^{-\alpha}$ divergence for the
half-filled homogeneous solution at the chemical potential  when $T\to
0$.  Since the norm of the spectral   
function is finite, the upper limit for $\alpha$ is one: $\alpha<1$. 
This new finding has not been previously discussed in the existing literature
on the $f$-spectral function of 
the FKM\cite{BrandtUrbanek92,FreericksZlatic2004}. 
The power-law scaling of the $f$-spectra is related to an
infra-red problem which --due to the logarithmic discretization-- can
properly be resolved within the NRG.
We will show that the exponent
$\alpha$ is proportional to the imaginary part of the bath field.  Since
small frequencies correspond 
to a very large time scale, this X-ray threshold behavior cannot be seen 
in the Keldysh-contour approach by integrating the equation of motion in real
time up to a finite cutoff in time\cite{BrandtUrbanek92,FreericksZlatic2004}.
For sufficiently high temperatures, where the X-ray singularities no longer
exist, our results are in good agreement with
Ref.~\onlinecite{FreericksZlatic2004}.  

Our investigation can also be motivated as a test or benchmark 
of the different possible
(numerical) many-body methods, which can be applied to the FKM (and thus to
correlated electron models, in general). To resolve fine structures on a
small frequency scale (around the chemical potential) the DMFT/NRG is
clearly superior to the Keldysh based method. Furthermore, the numerical
efforts and  resources needed are smaller by magnitudes; whereas
supercomputers had to be used in Ref.\onlinecite{FreericksZlatic2004}, our
results can be obtained on conventional PCs (or laptops) needing only
seconds of computation time. Finally, as a byproduct we obtain also the
$c$-electron Green function $G_c$ and self-energy $\Sigma_c$ within the
DMFT/NRG approach, which is known independently from the exact
self-energy functional. Though it is known that the NRG is slightly less
accurate in the reproduction of high-frequency 
features\cite{RaasUhrigAnders2003} (just because of the
logarithmic discretization and the resulting lower resolution 
at higher frequencies), very good agreement is obtained also for this
quantity.


\section{Theory}

\subsection{Dynamical Mean Field Theory}

The FKM was  solved exactly for $d \rightarrow \infty$ by Brandt
and Mielsch \cite{BrandtMielsch89}. Using the locality of the
self-energy $\Sigma_c$ \cite{MuellerHartmann89}, they showed that the
local conduction electron Green function for the homogeneous phase
must have the form 
\begin{eqnarray}
\label{eqn:gc-loc}
  G_c^{loc}(z) &=& \frac{1-n_f}{z +\mu -\e_c - \Gamma(z)} +\frac{n_f}{z+\mu -\e_c-\Gamma(z)-U} 
\end{eqnarray}
since the local $f$-electron number is conserved: $[H,f^\dagger_i
f_i]=0$. $n_f=\expect{f^\dagger_if_i}$ is the average number of $f$-electrons 
at site $i$, and
the complex dynamical field $\Gamma(z)$ contains the dynamics due to
the coupling of the effective site to the remainder of the system (lattice).
The local Green function
$G_c(z)$ can also be obtained from the self-energy $\Sigma_c$
\begin{eqnarray}
\Sigma_c(z) &= & z+\mu -\Gamma(z) - [ G_c(z)]^{-1}
\label{eqn:self-energy-c}
\end{eqnarray}
via Hilbert transformation
\begin{eqnarray}
  \label{eqn:gc-lattice-sum}  
  G_c^{latt}(z) &=& \int d\e \frac{\rho_0(\e)}{ z+\mu -\e - \Sigma_c }
\\
G_c^{latt}(z)&=& G_c^{loc}(z)
  \label{eqn:gc-latt=gc-loc}  
\punkt
\end{eqnarray}
For given $f$-occupation $n_f$,
Eqns.~(\ref{eqn:gc-loc}-\ref{eqn:gc-latt=gc-loc}) form a set of
self-consistency equations from which the dynamical field $\Gamma(z)$
and the self-energy $\Sigma_c(z)$ can be calculated. To this  end, 
the level position $\e_f$ has to be adjusted such that the average
$f$-occupation is given by $n_f$. For a fixed total particle number
$n_{tot}$  per site - $n_{tot} =n_f +n_c$ -
the chemical potential is adjusted accordingly.

\subsection{Dynamical Properties of the Effective Site}

The effective site can also be viewed as an effective single
impurity Anderson model (SIAM)\cite{Kim87,Jarrell92,Pruschke95,Georges96}
\begin{eqnarray}
  \label{eq:effecive-siam}
  H_{eff}&=& 
(\e_c-\mu) c^\dagger c + (\e_f -\mu) f^\dagger f
 \nonumber \\
&& +
\int d\e\; \e d^\dagger_{\e}  d_{\e}
+ U \hat n_c \hat n_f
 \nonumber \\
&& +
\int d\e\; V\sqrt{\rho_{\mbox{\tiny eff}}(\e)}
\left(
d^\dagger_{\e} c
+
c^\dagger d_{\e\sigma} 
\right)
\punkt
\end{eqnarray}
with an energy dependent hybridization function $\Delta(\e) =
V^2\rho_{\mbox{\tiny eff}}(\e) =\Im m \Gamma(\e-i0^+)$ describing the 
coupling of
the $c$-electron to a fictitious bath of ''conduction electrons'' created by
$d^\dagger$ with density of states (DOS) $\rho_{\mbox{\tiny eff}}(\e)$. 
The $f$-electron does not interact with the bath
field. The hybridization strength $V$ is chosen to be constant and
defined via 
\begin{eqnarray}
\pi  V^2  &=& \int d\e \, \Delta(\e)
\end{eqnarray}
and the DOS of the (fictitious) ($d$)
conduction electrons is given by
$\rho_{\mbox{\tiny eff}}(\e) =\Delta(\e)/(\pi V^2)$.

Using the equation of motion for Fermionic Green functions,
\begin{eqnarray}
  z \green{A}{B}(z) &=& \expect{\{A,B\} } + \green{[A,H]}{B}(z)
\end{eqnarray}
with the commutators $[c,H_{eff}]$ and $[f,H_{eff}]$,
it is straightforward to derive two 
exact relations for the local $c$- and $f$-Green function of the effective
site, $G_c(z)=\green{c}{c^\dagger}$ and $G_f(z)=\green{f}{f^\dagger}(z)$
\begin{eqnarray}
  \left(z-\e_c -\Gamma(z)\right) G_c(z) &=&1 + U \green{c f^\dagger
    f}{c^\dagger}(z) \\
  \left(z-\e_f\right) G_f(z) &=&1 + U \green{f c^\dagger
    c}{f^\dagger}(z)
\punkt
\end{eqnarray}
Parameterizing the Green functions via a self-energy $\Sigma_{c/f}$,
 $G_{c} = [z-\e_c -\Gamma(z)-\Sigma_c(z)]^{-1}$ and
 $G_{f} = [z-\e_f-\Sigma_f(z)]^{-1}$, yields
the exact relations
\begin{eqnarray}
\label{eqn:sigma-c}
\Sigma_c(z) &=& U \frac{F_c(z)}{G_c(z)}
\\ 
\label{eqn:sigma-f}
\Sigma_f(z) &=& U \frac{F_f(z)}{G_f(z)}
\end{eqnarray}
with $F_c(z) =   \green{c f^\dagger  f}{c^\dagger}$ and
$F_f(z) =   \green{f c^\dagger  c}{f^\dagger}$.
The problem to be solved is
 just a special case of a SIAM \cite{BullaHewsonPruschke98}
with the hybridization matrix element for one spin component set to
zero.

\subsection{The Numerical Renormalization Group (NRG)}
The Hamiltonian (\ref{eq:effecive-siam}) is solved using the
NRG\cite{Wilson75}. The core of the NRG approach is a logarithmic energy
discretization of the effective $d$-conduction band around the Fermi
energy, $\w_n^\pm = \pm D\lambda^{-n}$ and $\Lambda>1$,
and a unitary transformation of the basis $d_{\w_n^\pm\as}$ onto a basis
such that the Hamiltonian becomes tridiagonal. The first Fermionic
operator \cite{Wilson75} $b_{0\as}$ is defined as $b_{0}=\frac{1}{2}\sum_{n}
(d_{\w_n^+}+d_{\w_n^-})$. Only $b_{0}$
couples directly to the impurity degrees of freedom. 
Eq.~(\ref{eq:effecive-siam}) is recasted
as a double limit of a sequence of dimensionless NRG
Hamiltonians:
\begin{equation}
{\cal H} = \lim_{\Lambda \rightarrow 1^+}
           \lim_{N \rightarrow \infty}
           \left\{
                 D_{\Lambda} \Lambda^{-(N-1)/2} {\cal H}_{N}
           \right\} ,
\end{equation}
with $D_{\Lambda} = D(1 + \Lambda^{-1})/2$, and
\begin{eqnarray}
{\cal H}_N &=& \Lambda^{\frac{N-1}{2}}
               \left [
                \frac{(\e_f -\mu) f^\dagger f +(\e_c-\mu) c^\dagger
                  c}{D_{\Lambda}} 
             \right. 
\nonumber \\
&&
                + \left(
                      \tilde{V} b^{\dagger}_{0} c
                      + {\rm H.c.}
                 \right)
\nonumber \\
&&
 +              \left.
                     \sum_{n = 0}^{N-1}
                      \Lambda^{-\frac{n}{2}}
                     \xi_{n }
                     \left \{
                            b^{\dagger}_{n+1}
                            b_{n}
                            + {\rm H.c.}
                     \right \}
           +  
                     \sum_{n = 0}^{N}
                      \Lambda^{-\frac{n}{2}}
                     \eta_{n }b^{\dagger}_{n}b_{n}
               \right ] .
\nonumber
\end{eqnarray}
$\tilde{V}$ is  related to the hybridization width
$V= D_{\Lambda}\tilde{V} $
\begin{eqnarray}
  \pi V^2 = \int d\e \Delta(\e) 
\end{eqnarray}
while the pre-factor $\Lambda^{(N-1)/2}$
guarantees that the low-lying excitations of ${\cal H}_N$ are of order
one for all $N$ \cite{Wilson75}. The energy dependent hybridization
function $\Delta(\w)$ determines the coefficients $\xi_n$ and $\eta_n$
\cite{Wilson75,BullaPruschkeHewson97} by performing the unitary
transformation from the basis $d_{\w_n^\pm}$ to $b_n$.

The NRG is used to calculate the spectral function  of the Green
function  for the Fermionic operators $A$ and $B$ 
\begin{eqnarray}
\rho_{A,B}(\w) &=& \frac{1}{\pi} \Im m [G_{A,B}(\w -i0^+)  
\nonumber \\ &=&
\frac{1}{Z}
\sum_{n,m} 
\left( e^{ -\beta E_n } + e^{-\beta E_m} \right)  
A_{nm}B_{mn} 
\nonumber
\\
&& \times \delta(\w  + E_n -E_m)
\label{eqn_fermionic-lehmann}
\komma
\end{eqnarray}
where the matrix elements $A_{nm}=\matrixElement{n}{A}{m}$ are
calculated using an eigenbasis of the Hamiltonian $H$, $H\ket{n}=E_n\ket{n}$.
The spectral functions of $G_{c/f}$ and $F_{c/f}$
are obtained from their Lehmann representation
by broadening the $\delta$-function of the discrete spectrum on a
logarithmic scale, i.~e.~$\delta(\w-E)\to e^{-b^2/4}
e^{-(\log(\w/E)/b)^2}/(\sqrt{\pi} b|E|)$
~\cite{BullaHewsonPruschke98}. The broadening  
parameter is usually chosen as $0.5 < b <1$, and we used $b=0.8$.
In order to minimize the broadening error, the physical Green functions are
calculated indirectly using the  self-energies 
$\Sigma_{c/f}(z)$ given by Eqs. (\ref{eqn:sigma-c}) and
(\ref{eqn:sigma-f})~\cite{BullaHewsonPruschke98}. The self-energy
$\Sigma_c$ is redundant at this point since it was already obtained from the
self-consistency cycle
Eqs.~(\ref{eqn:gc-loc}-\ref{eqn:gc-latt=gc-loc}). However, it can be
used to quantify the discretization error of the NRG, especially at
high energies \cite{RaasUhrigAnders2003}.
Since the $f$-electrons do not hybridize, only a spinless bath must be
considered which makes the NRG calculations extremely fast.


\section{Results}

\begin{figure}[thb]
  \centering
  \includegraphics[width=85mm]{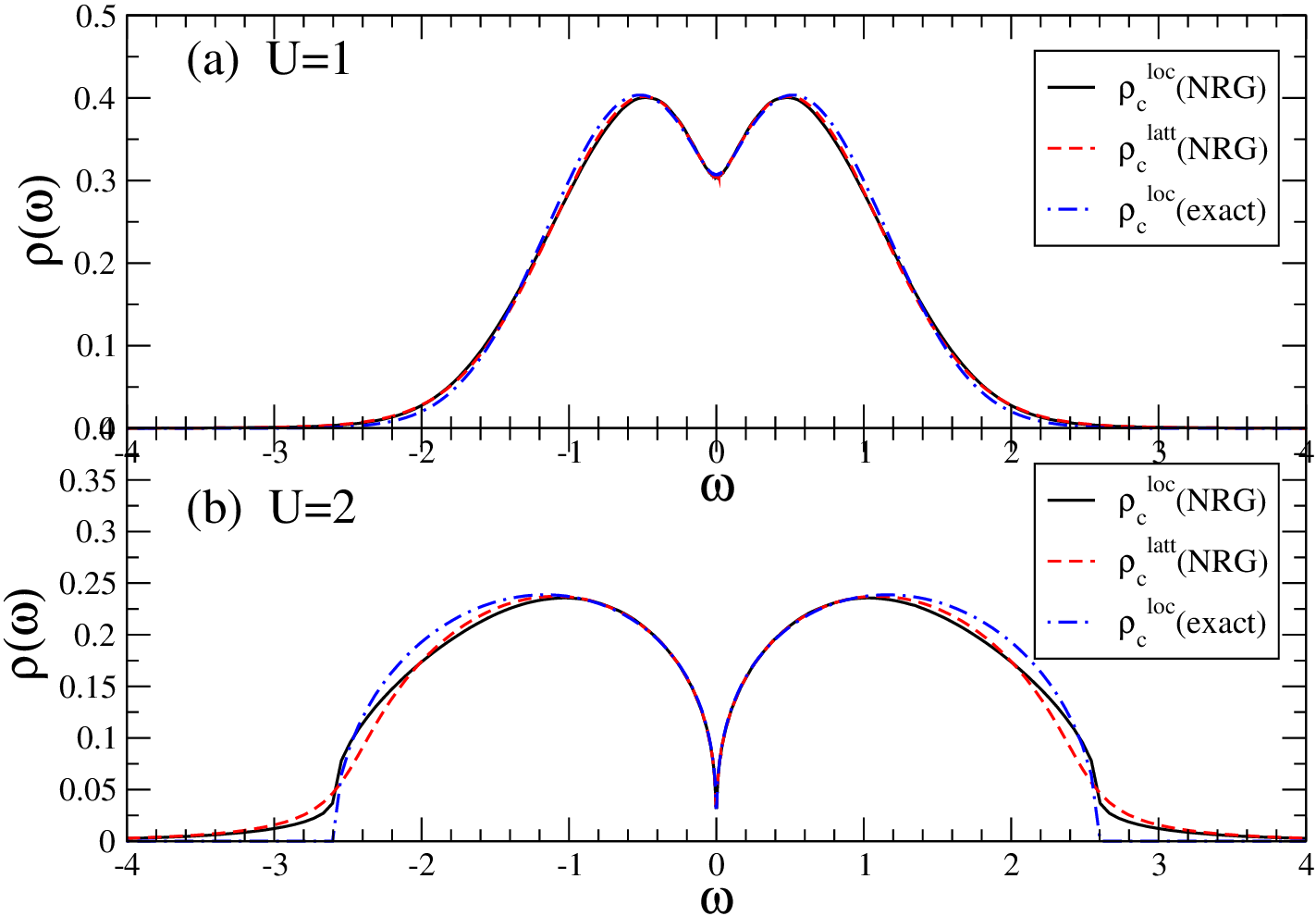}
  \caption{Spectral function of the $c$-Green function of the
    effective site (solid line), the lattice $c$-Green function,
    Eqn.~(\ref{eqn:gc-lattice-sum}), (dashed line) both calculated
    using the NRG, in comparison with the exact solution obtained from
  Eqns.~(\ref{eqn:gc-loc}-\ref{eqn:gc-latt=gc-loc}) (dotted-dashed
  line) for (a) a Gaussian DOS $\rho_0(\e)$ and U=1 and (b) a
  semi-circular DOS $\rho_0(\e)$ and $U=2$ for $T\to 0$. NRG parameters:
  $\Lambda=1.6$, $N=500$, $b=0.55$. }
\label{fig:gc-dos}
\end{figure}

\subsection{Spectral functions for $T\to 0$}

We investigate the FKM for the particle-hole symmetric
case in the absence of a symmetry breaking using two different densities of
states (DOS) for the unperturbed conduction electrons on the 
underlying lattice, namely a semi-circular DOS
$\rho_0(\e) = \sqrt{4-(\e/t^*)}/(2\pi t^2)$ (Bethe lattice) and a Gaussian DOS,
$\rho_0(\e) = \exp(-(\e/t^*)2)/(\sqrt{\pi}t^*)$ (hyper-cubic
lattice). The energy unit 
will be $t^*$ throughout the paper. At half filling, $n_{tot}=1$, and
for a given set of parameters $\mu=U/2, \e_c=\e_f=0$,  we solved the
Eqns.~(\ref{eqn:gc-loc}-\ref{eqn:gc-latt=gc-loc}) self-consistently to
determine the bath field $\Gamma(z)$.
Its imaginary part $\Delta(\w) = 
\Im m \Gamma(\w-i0^+)$ is used in the 
algorithm described in \cite{BullaPruschkeHewson97} to determine the
parameter  $\xi_n$ which enters the NRG of the effective site of the
problem; one has
$\eta_n=0$ in the particle-hole symmetric case for  all $n$.

As mentioned before, the self-energy $\Sigma_c$ provided by the NRG is
a redundant quantity. However, we obtain $G_c^{latt}(NRG)$ from
(\ref{eqn:gc-lattice-sum}) as well as $G_c^{loc}(NRG)$ via
$G_c^{loc}(NRG) = [z+\mu -\e_c -\Gamma(z) -\Sigma_c(z)]^{-1}$ 
(cf. (\ref{eqn:self-energy-c})) which we
compare with the exact result from the self-consistency equations
(\ref{eqn:gc-loc}-\ref{eqn:gc-latt=gc-loc}). Their spectral functions
are plotted in figure \ref{fig:gc-dos} for two different values of $U$
and two different DOS $\rho_0(\e)$. Fig.~\ref{fig:gc-dos}(a) shows a
comparison for $U=1$ using the Gaussian DOS $\rho_0$ while the
spectral functions for the semi-circular DOS and $U=2$ are plotted in
\ref{fig:gc-dos}(b). The agreement is very good. For $|\w|\to 0$, the
three spectral functions coincide within the numerical error, a sign
of a highly accurate calculation of the spectral function for low
frequencies due to the equation of motion. For large frequencies, the
NRG spectral functions tend to be broadened too much which also yields
an overestimation of the imaginary part of the self-energy
$\Sigma_c$. Therefore 
$\rho^{latt}_c(NRG)=\Im mG_c^{latt}(NRG)(\w-i0^+)/\pi$ turns
out to be slightly smaller than  $\rho^{loc}_c(NRG)$ for large
frequencies. This error is controlled by the NRG discretization
parameter  $\Lambda$ and can be reduced by reducing $\Lambda$ while
simultaneously increasing the number of states $N_s$ kept after each
NRG iteration. For our calculation, we used $N_s=500$ and
$\Lambda=1.6$. We like to emphasize that we only need to add a
spin-less chain link in each NRG iteration which increases the number
of states by a factor of $2$ contrary to the usual single
impurity Anderson or Kondo model where $4$ states are needed to
describe a new chain link which makes the NRG calculations extremely
fast as mentioned before.

Two discretization errors determine the
deviation of the high energy curves: the discretization errors in the
determination of the NRG hopping parameters $\xi_n$ by averaging
$\Delta(\w)$ over the energy intervals
$D[\Lambda^{-n},\Lambda^{-n+1}]$ and the artificial broadening of the
$\delta$-function in the Lehmann representation of the spectral function
(\ref{eqn_fermionic-lehmann}). It is know that the latter is partially
compensated using the equation of motion\cite{BullaHewsonPruschke98}.
These systematic errors occur in all applications of the NRG as solver
of the effective single-site problem in the dynamical mean field
theory. However, these errors do not have any influence on the
spectral properties for $|\w|\to 0$. Here, the NRG furnishes an
unprecedented resolution due to the logarithmic discretization of the
energy mesh. This enables us to reveal the X-ray properties of the
$f$-spectral function not visible 
in the previously published
calculations\cite{BrandtUrbanek92,FreericksZlatic2004}, because the --in
principle exact-- method used there requires a discretization and
time cutoff in the numerical evaluation limiting the accessible resolution.

\begin{figure}[htbp]
  \centering

 (a) \includegraphics[width=85mm,clip=true]{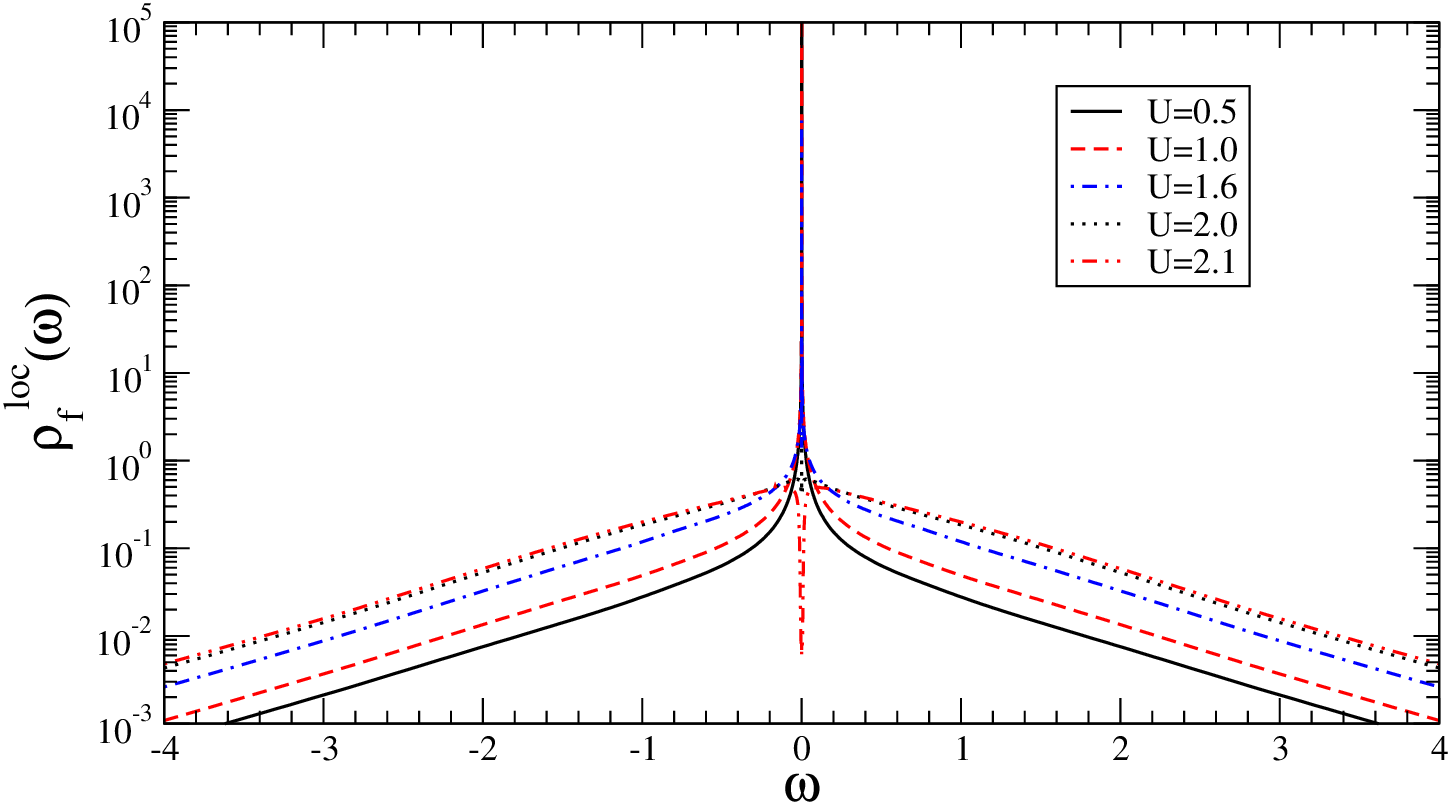}\\
\vspace*{10mm}

 (b) \includegraphics[width=85mm,clip=true]{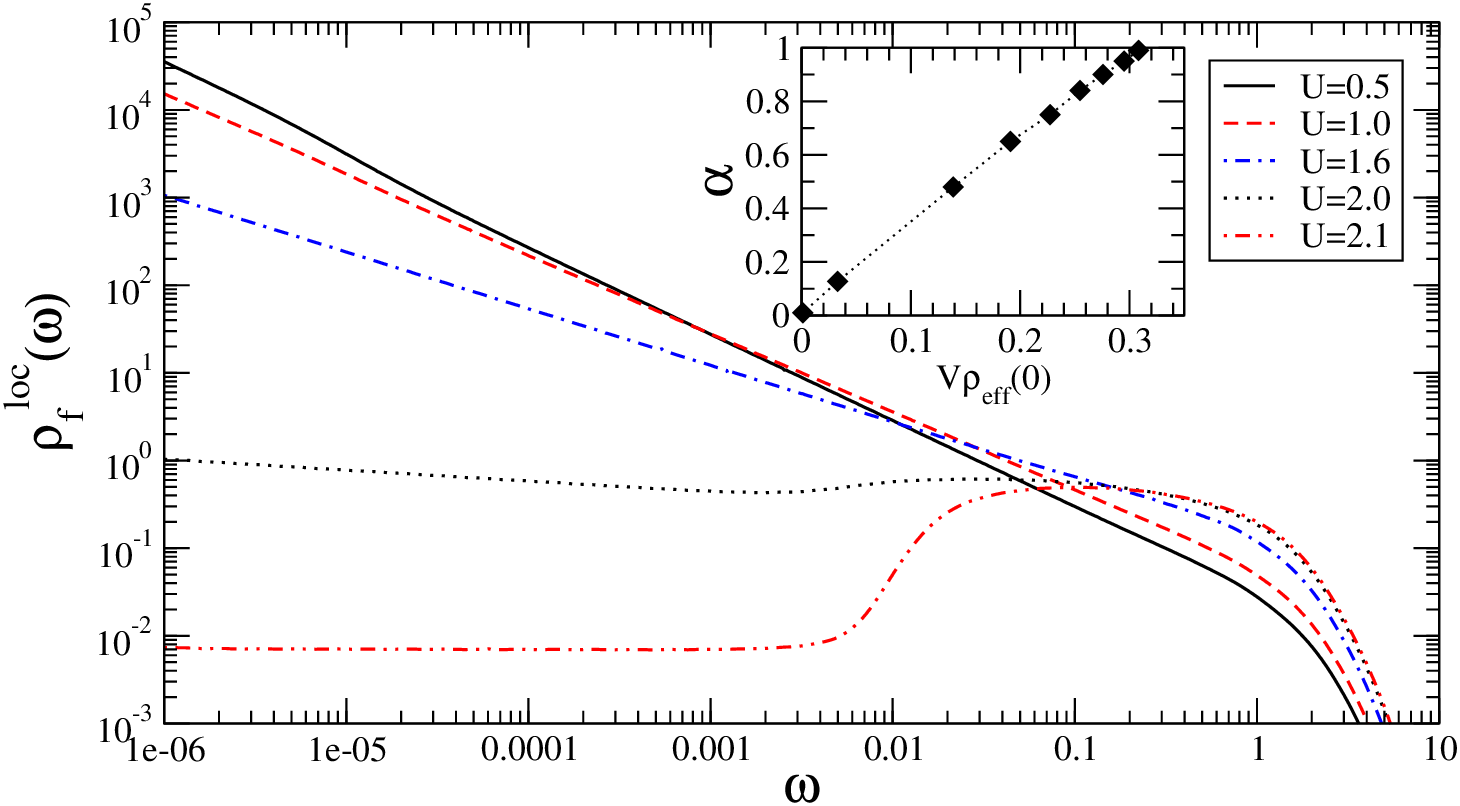}

  \caption{$f$-spectral function for different $U$ for the
    semi-circular $\rho_0(\e)$.
    (a) shows the spectra on a linear-log scale, (b) the same data on a
    log-log scale for positive frequencies. For $|\w| \to 0$, the
    spectral function has an algebraic singularity: $\rho_f(\w)
    \propto |\w|^{-\alpha}$. The inset of (b) displays the linear
    dependence of $\alpha$  on the  dimensionless
    coupling constant $g=V\rho_{\mbox{\tiny eff}}(0)$ for the metallic
    case $g >0$.
    With    increasing $U$, $g$ decreases monotonically.
    NRG parameters as in
    Fig.~\ref{fig:gc-dos}
  }
  \label{fig:gf-semi-circular-t0}
\end{figure}

After we have established the excellent agreement between the NRG
solution for the FKM and exact result for the $c$-electron spectral
function, we will focus on the $f$-spectral function in the remainder  of
the paper. The $f$-spectral function for the FKM with semi-elliptical
DOS is shown in Fig.~\ref{fig:gf-semi-circular-t0} for $T\to
0$. Clearly visible in the linear-log plot
\ref{fig:gf-semi-circular-t0}(a) is the divergence of the spectral
function for $|\w|\to 0$, while Fig.~\ref{fig:gf-semi-circular-t0}(b)
reveals its algebraic nature for the metallic case ($g> 0$):
$\rho_f(\w)  \propto 
|\w|^{-\alpha}$. The $f$-Green function describes the response of the
system to the addition or removal of an $f$-electron onto a lattice
site. This process suddenly changes the Coulomb potential for the itinerant
conduction electrons. Since the $f$-electron number is conserved in
the absence of a $t_{ff}$-hopping matrix element, the bath conduction
electrons have to relax to the new ground state under the presence of
the time independent additional scattering potential in a similar way
as in the X-ray edge problem \cite{RouletGavNozieres69}. This infinite
cascade of particle-hole excitations generates a logarithmic slow-down
which yields an algebraic singularity in the spectral function
$\rho_f(\w)$. The connection between these two problems is
mentioned in 
Refs.\onlinecite{SiKotliarGeorges1992,MoellerRuckensteinSchmittRink1992},
for instance. But previous
calculations of the  $f$-spectral functions
\cite{BrandtUrbanek92,FreericksZlatic2004} do not reveal this feature
since the method used there is based on a real time integration 
along Keldysh contours. In order to see the algebraic singularities,
numerical integrations at $T=0$ to exponentially large time scales would
have to be performed, which is not possible in practice also because of the
possible accumulation of numerical errors.

The inset of  Fig.~\ref{fig:gf-semi-circular-t0}(b) shows the
dependence of the exponent $\alpha$ on the dimensionless coupling
constant $g=V \rho_{\mbox{\tiny eff}}(0) = 
\Im m\Gamma(0-i0^+)/(V\pi)$ at the
chemical potential. With increasing $U$, $g$ is monotonically
decreasing. At the same time, the exponent $\alpha$ is also reduced as can
be seen in Fig.~\ref{fig:gf-semi-circular-t0}(b). We find a linear
connection between $g$, which is implicitly $U$ dependent, and the
exponent $\alpha$. A relation between $U$ and $\alpha$ was already 
discussed in Ref.\onlinecite{SiKotliarGeorges1992}. There --for the
(unphysical) assumption of a Lorentzian  unperturbed DOS-- a mapping on 
Nozieres' original X-ray problem\cite{RouletGavNozieres69,Nozieres71} was
made. But obviously the exponent $\alpha$ obtained within our DMFT/NRG
treatment approaches $0$ at a critical interaction strength $U_c$, at which
the metal-insulator transition sets in, whereas no critical $U_c$ is
obtained in Ref.\onlinecite{SiKotliarGeorges1992}, probably because only a
single-impurity (and no selfconsitent DMFT-scheme) is considered there.
Since the total spectral weight of $\rho_f$ is 1, the integral over a 
small interval around zero, on which the power law behavior is valid,
must be finite, too,   i.~e. 
\begin{eqnarray}
\int_{-\eta}^{\eta} d\w \rho_f(\w) &=&
a\int_{-\eta}^{\eta} d\w |\w|^{-\alpha} < 1
\label{eqn:spectral-weight}
\end{eqnarray}
Therefore, $\alpha<1$ must hold.
The exponent $\alpha$ obtained by a   power-law fit to the spectral
function up to $|\w|\approx 10^{-8}$ is consistent with this
analytical requirement.  

\begin{figure}[htbp]
  \centering

 (a) \includegraphics[width=85mm,clip=true]{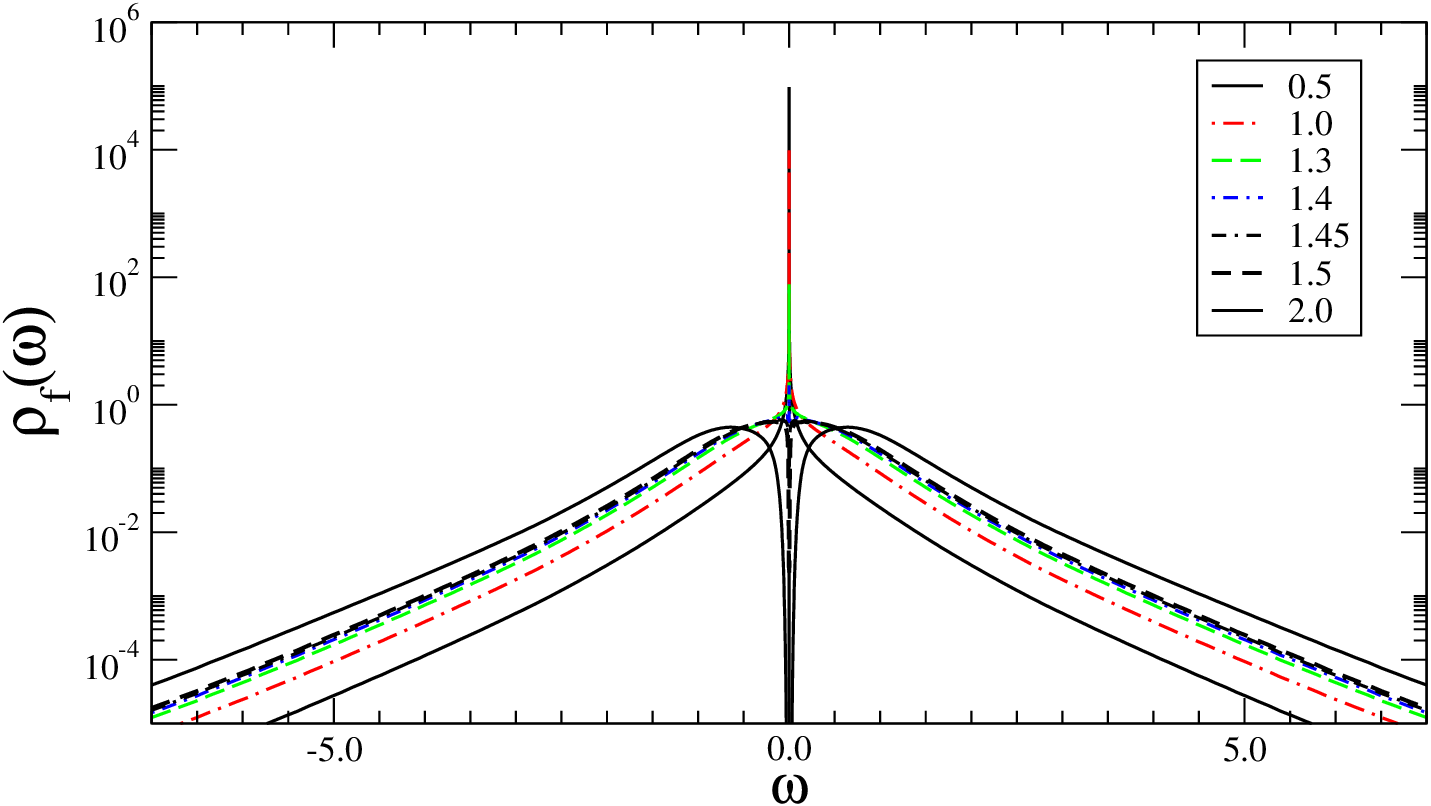}\\
\vspace*{10mm}

 (b) \includegraphics[width=85mm,clip=true]{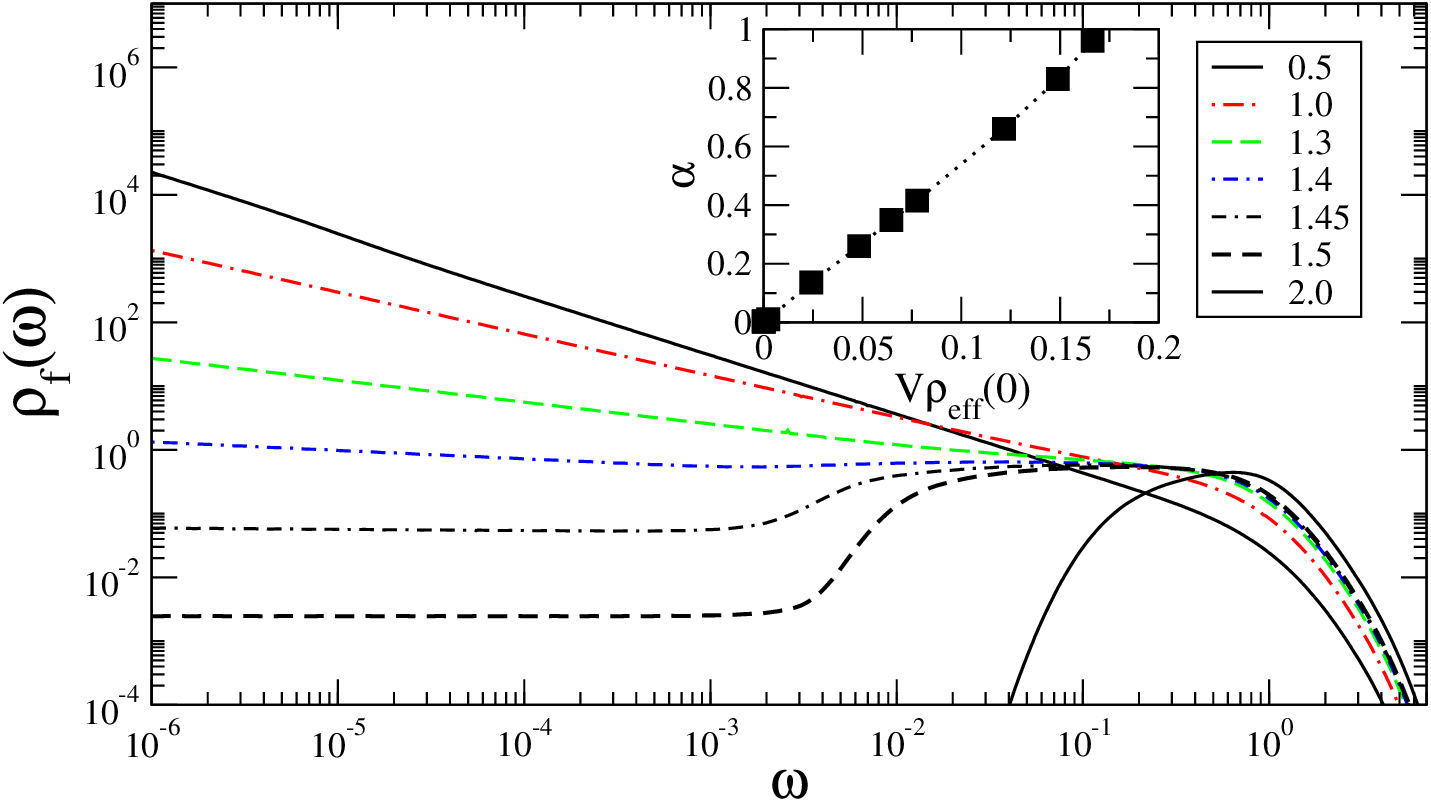}

  \caption{$f$-spectral function for different $U$ for the
        Gaussian DOS $\rho_0(\e)$.
    (a) shows the spectra on a linear log scale, (b) the same data on a
    log-log scale for positive frequencies. For $|\w| \to 0$, the
    spectral function has an algebraic singularity: $\rho_f(\w)
    \propto |\w|^{-\alpha}$. The inset of (b) displays the linear
    dependence of $\alpha$ on the  dimensionless
    coupling constant $g=V\rho_{\mbox{\tiny eff}}(0)$ of the effective site.
    NRG parameters as in
    Fig.~\ref{fig:gc-dos}
  }
  \label{fig:gf-gauss-t0}
\end{figure}

Figure \ref{fig:gf-gauss-t0} shows the evolution of the $f$-spectral
function $\rho_f(\w)$ for a Gaussian unperturbed DOS $\rho_0(\w)$ 
as function of
$U$ for the same NRG parameter as in
Fig.~\ref{fig:gf-semi-circular-t0}. In the chosen units, the
square of the effective hybridization strength, $V^2=0.5$ is half the
size of the value for $V^2$ for the semi-circular DOS. Therefore, the
metal-insulator transition occurs at a smaller  value of $U$. Setting
aside this difference and the slightly different shapes, the spectra are
similar. As expected, the spectral functions also show a power law
behavior for $|\w|\to 0$.

We have also checked the basic sum rules $\rho_f$ has to
fulfill\cite{FreericksZlatic2004}. Whereas the basic rules $\int d\w
\rho_f(\w) = 1 , \int d\w \rho_f(\w) f(\w) = 0.5$ and $\int d\w \rho_f(\w)\w
= 0$ are exactly fulfilled, the rule\cite{FreericksZlatic2004} $\int d\w
\rho_f(\w)\w^2 = U^2/4$ 
is slightly violated within our  NRG-scheme. We
obtain a proportionality to $U^2$ but with a larger prefactor, probably as
the NRG is less accurate at higher frequencies which get a  higher weight
due to the $\w^2$-factor under the integral.

\subsection{$f$-Electron Self-energy}

\begin{figure}[htbp]
  \centering

 \includegraphics[width=80mm,clip=true]{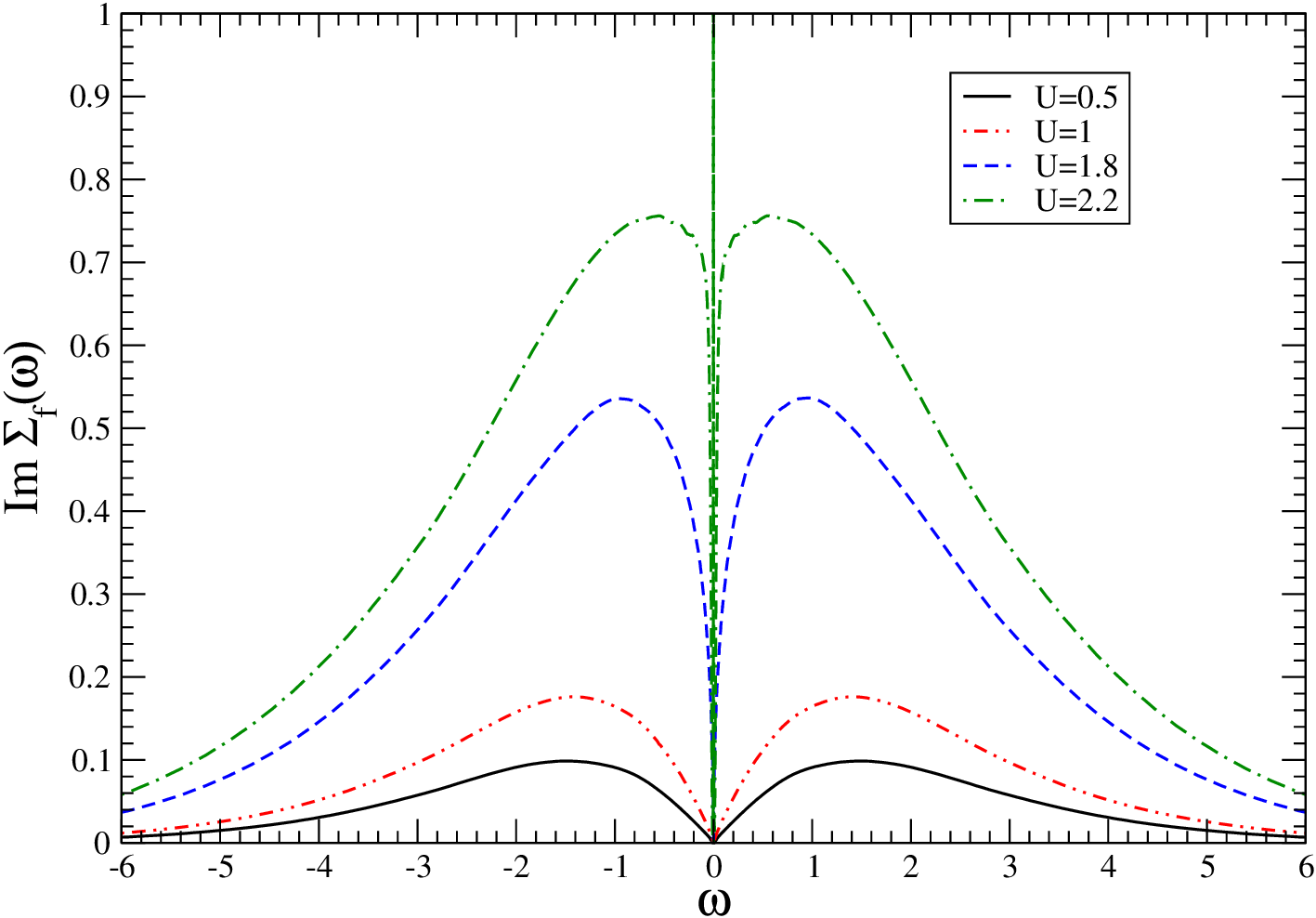}

  \caption{Imaginary part of the $f$-electron self-energy $\Im m
    \Sigma(\w-i0^+)$ for different values of $U$ and a
semi-circular unperturbed DOS. The scattering 
rate increases with increasing $U$ as expected from
Eqn.~(\ref{eqn:sigma-f}). 
        Parameters as in
    Fig.~\ref{fig:gf-semi-circular-t0}.
  }
  \label{fig:imag-self-f}
\end{figure}


The imaginary part of 
$\Sigma_f(\w-i0^+)$ is shown in Fig.~\ref{fig:imag-self-f} for 4
different values of $U$ and a semi-circular unperturbed DOS. 
For the metallic case, $U=0.5,1,1.8<U_c$, 
$ \Im m \Sigma_f(\w-i0^+) \to 0$ for $\w\to 0$ as expected from the
suppression of scattering processes on conduction electrons. However,
obviously we observe here again a non-analytic, power-law behavior.
Therefore, the $f$-electron spectral function and self-energy of the FKM 
provides for an example and paradigm of a system exhibiting non Fermi liquid
behavior. As it is well known, for a normal Fermi liquid the self-energy
imaginary part has to vanish according to an $\w^2$-behavior at the chemical
potential. But here we have obviously no $\w^2$-behavior but a non-analytic
vanishing of $\Im m \Sigma_f(\w)$. Via  
$G_{f} = [z-\e_f-\Sigma_f(z)]^{-1}$ this non-analytic self-energy behavior 
translates
to a divergence in the spectral function for $|\w|\to 0$. Above a critical
value of the Coulomb interaction $U$, namely for 
$U=2.2>U_c$, the solution of the DMFT yields an insulating behavior, i.e. a
gap in the $f$-electron spectral function.  Then, the
imaginary part of $\Sigma_f(\w-i0^+)$ has an additional
$\delta(\w)$ peak, i.e. there $\Sigma_f(\omega-i0^+)$ has a pole at
$\w=0$, which leads to a vanishing $f$-spectral function in the
energy-gap around the chemical potential. This metal insulator transition
observed here is similar to the Mott-Hubbard transition occuring for the
Hubbard model and for the $c$-electron spectral function of the FKM.

\subsection{Finite Temperature}

\begin{figure}[htbp]
  \centering

 \includegraphics[width=87mm,clip=true]{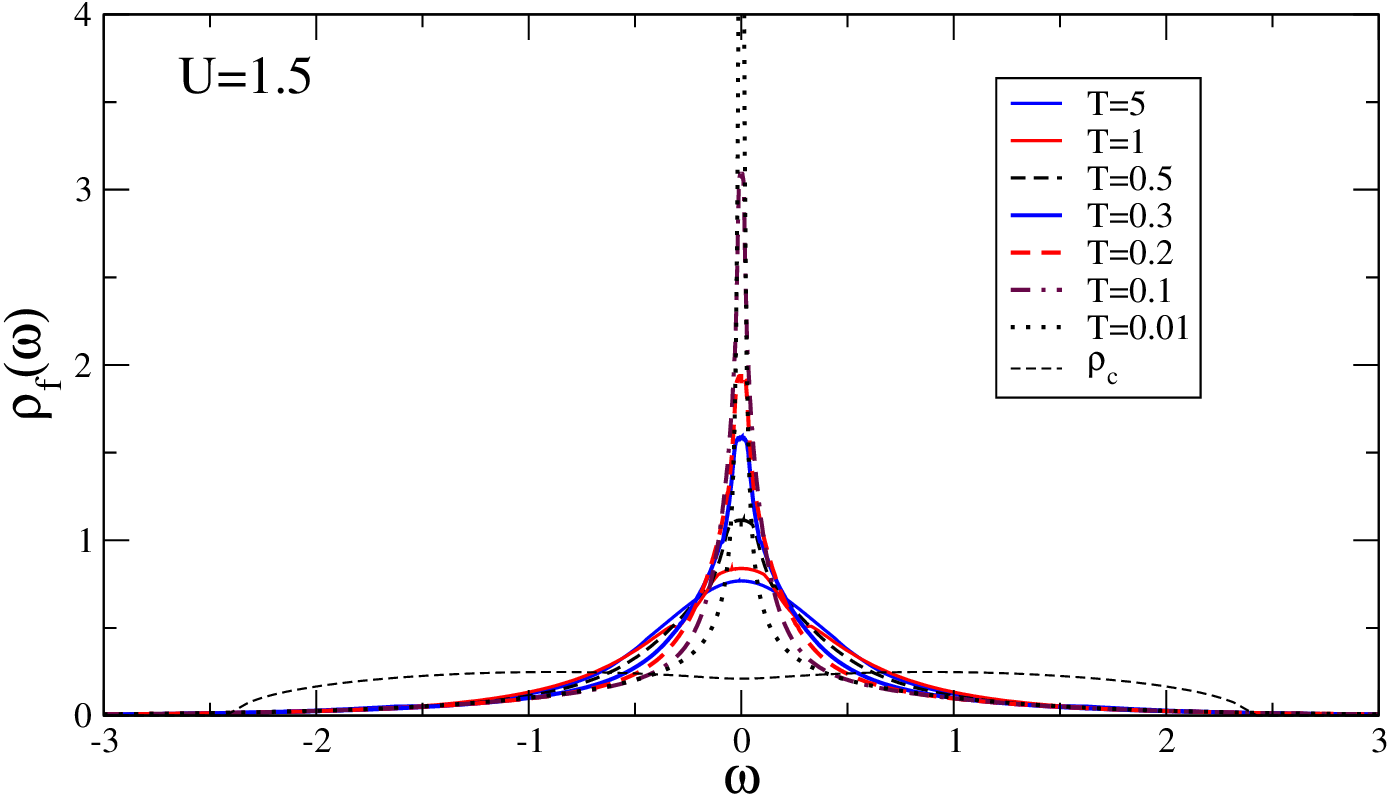}

  \caption{Evolution of the finite temperature spectral function
    $\rho_f(\w)$ for a semi-circular DOS in the metallic phase for
    $U=1.5$. We truncated the $\rho(\w)$-axis in favor of the high
    temperature spectral information. While for
  high temperatures, the peak at $\w=0$ is rather wide, the
  development of the singularity is clearly visible for the smallest
  temperature plotted here, $T=0.01$.     NRG parameters as in
    Fig.~\ref{fig:gc-dos}.
  }
  \label{fig:finite-temp-metal}
\end{figure}

\begin{figure}[htbp]
  \centering

  \includegraphics[width=87mm,clip=true]{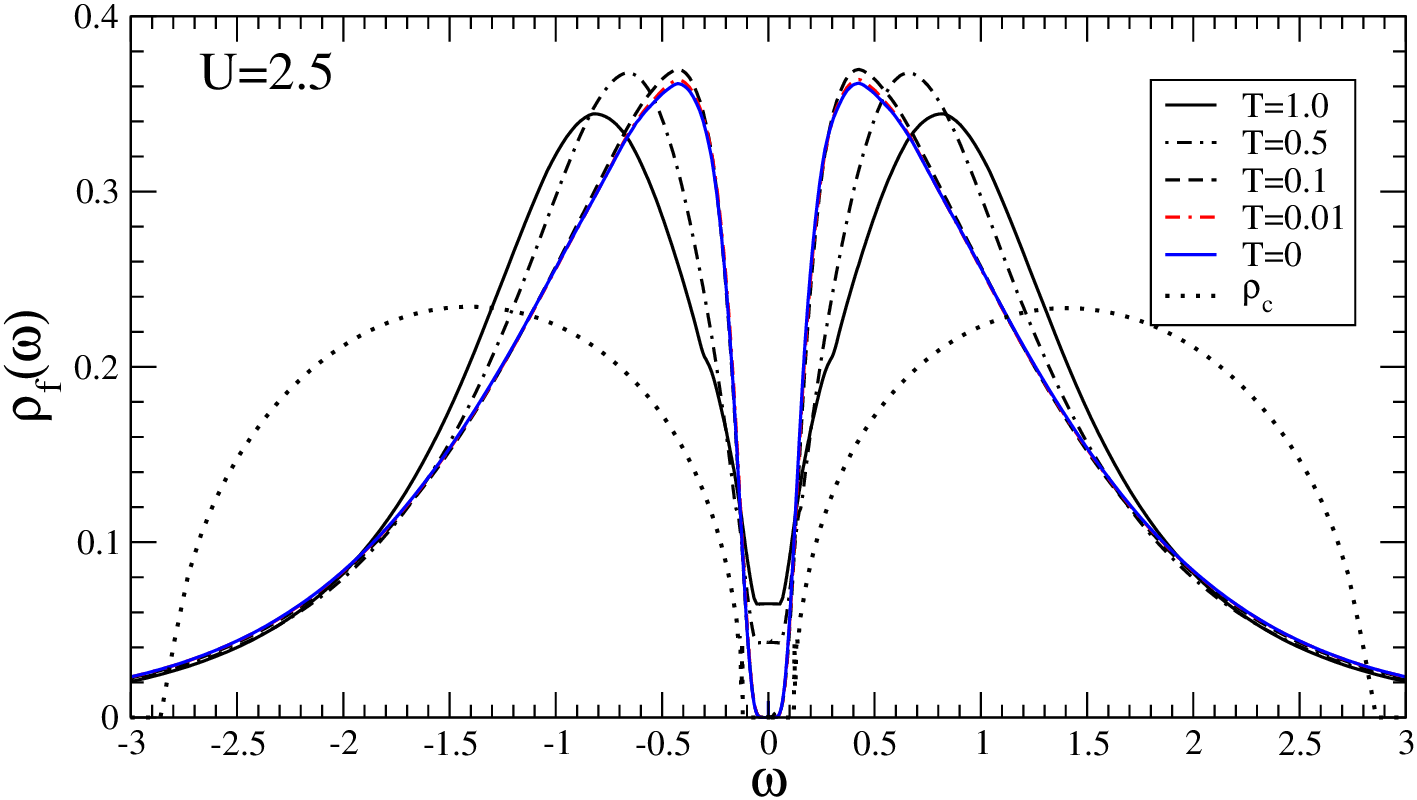}
  
  \caption{Evolution of the finite temperature spectral function
    $\rho_f(\w)$ for a semi-circular DOS in the insulating  phase for
    $U=2.5$. While for
  high temperatures spectral weight is still found in the gap between
  the two sub-bands, it is rapidly vanishing for $T<0.1$. NRG parameters as in
    Fig.~\ref{fig:gc-dos}.
  }

  \label{fig:finite-temp-insulator}
\end{figure}

We used a newly developed finite temperature algorithm, as described in
appendix \ref{sec:nrg-gf-FT}, to calculate the NRG spectral functions
which enter the Eqns.~for the local self-energies, (\ref{eqn:sigma-c})
and (\ref{eqn:sigma-f}). Figure \ref{fig:finite-temp-metal} shows the
$f$-spectral function for different temperatures in the metallic phase
for $U=1.5$. At  high temperatures, the peak at $\w=0$ is rather
broad. Lowering the temperature, the algebraic singularity
clearly gradually develops. 
Since we have established its existence in the normal
metallic phase in the previous section, we truncated the height of the
$\rho(\w)$-axis in favor of the high temperature spectral
information. These results agree nicely with those obtained recently  by
Freericks et.~al.~\cite{FreericksZlatic2004}, compare their Fig.~6. 
Since they obtain their
spectral function at finite temperature by integration in the time
domain, their spectral function is resolution limited by  the
largest time $t_l$ of the numerical integration, $\w_{min}=1/t_l$. For
$T\to 0$, they extrapolate $\rho_f(0)$ to a finite value, which is
essentially displayed in the inset in Fig.~6 of
Ref.~\onlinecite{FreericksZlatic2004}. However, our work shows that
$\rho_f(\w=0,T\to 0)\to \infty$ with a power law $\rho_f(\w) \propto
|\w|^{-\alpha}$ which could not be seen in 
Ref.~\onlinecite{FreericksZlatic2004}
because of the limited frequency resolution inherent to the method used.
Figure \ref{fig:finite-temp-insulator} displays the evolution of the
finite temperature $f$-spectral function in the  insulating
phase for $U=2.5$. While at higher temperature, finite spectral weight
is found in the vicinity of $\w=0$, rather rapidly a gap
develops for temperatures $T<0.1$. The difference between the
$T=10^{-2}$ and the $T=0$ spectral function is marginal for $|\w|> T=10^{-2}$.
 This behavior is again essentially in agreement with results obtained in
Ref.~\onlinecite{FreericksZlatic2004}, compare Fig.~7 in that reference.


\section{Discussion and Conclusion:}

We calculated the f-electron spectral function $\rho_f$ 
of the FKM at half filling in
the $d \rightarrow \infty$ limit using the DMFT/NRG method. $\rho_f$ is
strongly temperature dependent, for small values of $U$, 
it has a single peak structure with a
power law singularity $|\w|^{-\alpha}$ for $\w \to 0$ at zero 
temperature $T=0$. For larger $U > U_c$,
a metal-insulator (Mott-Hubbard) transition
occurs and $\rho_f$ gets a two-peak structure with a gap at $T=0$.
Therefore, the obvious connection of the FKM with the X-ray threshold
problem \cite{SiKotliarGeorges1992,MoellerRuckensteinSchmittRink1992}
has been explicitely worked out and demonstrated here. 
The  imaginary part of the f-electron self-energy vanishes for
$\w\rightarrow 0$ for  small $U < U_c$, but it does not follow the
standard Fermi liquid $\w^2$-law, but it vanishes non-analytically,
i.e.~one obtains non Fermi liquid behavior
\cite{SiKotliarGeorges1992}, which leads to the power 
law singularity in $\rho_f$. Above the metal-insulator transition,
$\alpha<0$ and the threshhold $E_0 =f(U-U_c)$ being a function of
$U-U_c$ is shifted to a finite energy. At finite temperatures,
precursors of the algebraic signularities are found which are cut of
by thermal fluctuations below $|\w|<T$.
Furthermore, the present investigation demonstrates the strength and
accuracy of the DMFT/NRG method. It is not only able to resolve low
frequency or temperature features with unprecedented accuracy, it is also
very efficient numerically requiring only minimal resources and computation
time; typical run-times have been around 2 minutes on a Pentium M laptop.
 
\begin{acknowledgments}

One of us (G.C.) thanks Jim Freericks, Veljko Zlatic and Romek Lemanski for
numerous, very useful discussions on the FKM and for 
informations on their recent work on that model, and we thank Jim Freericks
and Veljko Zlatic for usefull comments on a first preprint version of this
manuscript.
\end{acknowledgments}

\appendix

\section{Calculation of the Finite Temperature Green Function using
  the NRG}

\label{sec:nrg-gf-FT}

We used a slight modification of the algorithm for finite temperature
Green functions by Bulla et.~al.\cite{BullaCostiVollhardt01}. Our 
algorithm has originally been developed for multi-band models.
In this case, collecting all matrix elements which contribute to
$\rho_{A,B}$ (\ref{eqn_fermionic-lehmann}) at each NRG iteration and
patching them together as described in Fig.~2 of
Ref.~\onlinecite{BullaCostiVollhardt01} is not practical in the two-band model:
the number of matrix elements exceeds rather rapidly even the memory 
of modern supercomputers for accurate calculations. The patching of
information from different iterations, however, is a very useful
concept since (i) it increases the accuracy in combination with  (ii)
producing additional pole positions by shifting the eigenenergies
slightly through iteration. Latter helps to mimic  a continuum
of states.

Instead of averaging individual matrix elements for 
different iterations, we average different $\rho_{A,B}^M(\w)$ for
fixed frequency $\w$ but different iterations $M$. The basic idea is
to combine the method of Costi et.~al.\cite{CostiHewsonZlatic94} with the
algorithm of Bulla  et.~al.\cite{BullaCostiVollhardt01}. The eigenstates of
iteration $M$ represent excitations on an energy scale $\w_M$
\begin{eqnarray}
  \label{eq:nrg-wn}
  \w_M &=& D\frac{1+\Lambda^{-1}}{2} \Lambda^{-M/2}
\end{eqnarray}
For each iteration $M$, we define a logarithmic grid interval $I_M$
for $3L+1$ frequencies
\begin{widetext}
\begin{eqnarray}
 I_M&=& \w_M\times
\left[
\Lambda^{-1},\Lambda^{-(L-1)/L},\cdots,
\Lambda_i=\Lambda^{(i-L)/L}
\cdots,\Lambda,\cdots,\Lambda^2
\right] 
\end{eqnarray}  
\end{widetext}
and evaluate $\rho_{A,B}^M(\w_M\Lambda_i)$
via (\ref{eqn_fermionic-lehmann}) for a fixed
iteration $M$. Since the information between even and odd iterations
are collected separately, $I_M$ overlaps with the $I_{M-2}$, i.e.~$I_M
\cap I_{M-1} = \w_M\times \left[1,\cdots,\Lambda^2\right]$ and
$I_{M}\cap I_{M-4} 
=\w_M\times \left[\Lambda,\cdots,\Lambda^2\right]$. 
Only the $\rho_{A,B}^M(\w)$ evaluated at the frequencies
$I_M'=\w_M\times\left[1,\cdots,\Lambda\right]$  provide accurate
information. At each iteration, we add the new data points
$\rho_{A,B}^M(\w_i)$ at $\w_i(M) =\w_M \Lambda^{(i-L)/L}$,
$i\in[0,3L]$ to the previously obtained set of spectral information
by
\begin{widetext}
\begin{eqnarray}
  \rho'_{A,B}(\w_i(M)) &=&
\left\{
  \begin{array}{l}
\displaystyle \rho_{A,B}(\w_i(M))\frac{\w_{i}-\w_{L}}{\w_{3L}-\w_{L}}
+
 \rho^M_{A,B}(\w_i(M))\frac{\w_{3L}-\w_{i}}{\w_{3L}-\w_{L}}
\hspace{3mm} \mbox{for} \;\; i\ge L
\\
 \rho^M_{A,B}(\w_i(M))
\hspace{3mm} \mbox{for} \;\; i< L
  \end{array}
\right.
\punkt
\end{eqnarray}
\end{widetext}
We do the same for the spectral information at the negative
frequencies $-\w_i(M)$. With this recursion relation, we mimic the
patching of the residua since residua far away from the frequency $\pm
w_i(M)$ will only contribute marginal through the broadening procedure. 
It turned out that $L=3-7$ is a good choice, and for this work, we
used $L=6$. Since the spectral information $ \rho^N_{A,B}(\w_i(N))$
for $i=0,\ldots,L-1$ is not very reliable for the last NRG iteration $N$, we
drop these $L$ frequencies.

The major advantage of this new algorithm is that any temperature
dependent spectral function can be calculated on the fly. There is no
need for storage of information of previous iterations other than the
spectral function $\rho_{A,B}(\w)$ itself. We also use two
different broadening functions 
\begin{widetext}
\begin{eqnarray}
  \label{eq:broading-functions}
\delta(\w-E) &\to &
  \left\{
  \begin{array}{lcr}
\displaystyle e^{-b^2/4} e^{-(\log(\w/E)/b)^2}/(\sqrt{\pi} b|E|)
&\mbox{for} & |E|\ge L_w T
\\
\displaystyle
\frac{1}{\pi} \frac{L_T T}{(\w-E)^2 + (L_T T)^2}
&\mbox{for} & |E|<L_w T
  \end{array}
\right.
\end{eqnarray}
\end{widetext}
depending whether the pole position $E$ is larger than a multiple of the
temperature $T$, $E \ge L_w T$ or $E < L_w T$. $L_w$ is a constant of
the order of $1$ as suggested in the literature
\cite{BullaCostiVollhardt01}. The parameter $L_T$ 
controls the width of the Lorentzian on an energy scale of $T$.
Here, we used $L_w=0.3$ and $L_T=0.1$. Since the iteration $M$ also
represents a characteristic temperature scale \cite{Wilson75}, we
only include spectral information up to iterations
\cite{CostiHewsonZlatic94} for which $\w_M>
L_{iter} T$ , where we chose $L_{iter}=0.1$.


\begin{thebibliography}{10}

\bibitem{FalicovKimball69}
L.~M. Falicov and J.~C. Kimball, Phys.~Rev.~Lett. {\bf 22},  997  (1969).

\bibitem{Hubbard64}
J. Hubbard, Phys. Roy. Soc. A {\bf 281},  401  (1964).

\bibitem{KennedyLieb86} 
T. Kennedy, E. Lieb, Physica A {\bf 138}, 320 (1986)

\bibitem{FreericksZlatic03} J. Freericks, V. Zlatic, Rev. Mod. Phys. {\bf
75}, 1333 (2003)


\bibitem{BrandtMielsch89}
U. Brandt and C. Mielsch, Z. Phys. B {\bf 75},  365  (1989).

\bibitem{MetznerVollhardt89}
W. Metzner, D.Vollhardt, Phys. Rev. Lett. {\bf 62}, 324 (1989)



\bibitem{BrandtUrbanek92}
U. Brandt and M.~P. Urbanek, Z. Phys. B {\bf 22},  297  (1992).

\bibitem{FreericksZlatic2004}
J.~K. Freericks, V.~M. Turkowski, and V. Zlatic, preprint, 
cond-mat  0407411  (2004).


\bibitem{SiKotliarGeorges1992} 
Q.~Si, G.~Kotilar and A.~Georges,  Phys. Rev. B, {\bf 46}, 1261 (1992)

\bibitem{MoellerRuckensteinSchmittRink1992}
G.~M\"oller, A.~Ruckenstein and S.~Schmitt-Rink, 
Phys. Rev. B, {\bf 46}, 7427 (1992)

\bibitem{Georges96}
A. Georges, G. Kotliar, W. Krauth, and M.~J. Rozenberg, Rev. Mod. Phys. {\bf
  68},  13  (1996)

\bibitem{Wilson75}
K.~G. Wilson, Rev. Mod. Phys. {\bf 47},  773  (1975).

\bibitem{RaasUhrigAnders2003}
C. Raas, G.~S. Uhrig, and F.~B. Anders, Phys.~Rev.~B {\bf 69},  041102(R)
  (2003).

\bibitem{MuellerHartmann89}
E. M\"uller-Hartmann, Z. Phys. B. {\bf 76},  211  (1989).

\bibitem{Kim87}
C.~I. Kim, Y. Kuramoto, and T. Kasuya, Solid State Commun. {\bf 62},  627
  (1987).

\bibitem{Jarrell92}
M. Jarrell, Phys.~Rev.~Lett. {\bf 69},  168  (1992).

\bibitem{Pruschke95}
T. Pruschke, M. Jarrell, and J.~K. Freericks, Adv. in Phys. {\bf 42},  187
  (1995).


\bibitem{BullaHewsonPruschke98}
R. Bulla, A.~C. Hewson, and T. Pruschke, J. Phys.: Condens. Matter {\bf 10},
  8365  (1998).

\bibitem{BullaPruschkeHewson97}
R. Bulla, T. Pruschke, and A.~C. Hewson, J. Phys.: Condens. Matter {\bf 9},
  10463  (1997).


\bibitem{RouletGavNozieres69}
B. Roulet, J. Gavoret, and P. Nozieres, Phys. Rev. {\bf 178},  1072  (1969).

\bibitem{BullaCostiVollhardt01}
R. Bulla, T.~A. Costi, and D. Vollhardt, Phys.~Rev.~B {\bf 64},  045103
  (2001).

\bibitem{CostiHewsonZlatic94}
T. Costi, A.~C. Hewson, and V. Zlatic, J. Phys.: Condens. Matter {\bf 6},  2519
   (1994).

\bibitem{Nozieres71} M. Combescot, P. Nozieres, J. Physique (Paris) {\bf
32}, 913 (1971)

\end{thebibliography}

%

\end{document}